\documentclass{aa}
\usepackage{graphicx}
\begin{document}
   \title{The Magnetic Field Structure in W51A }


   \author{Antonio Chrysostomou\inst{1}, David K. Aitken\inst{1}, Tim Jenness\inst{2}, Christopher J. Davis\inst{2}, \\
	 J.H. Hough\inst{1}, Rachel Curran\inst{1} \and M. Tamura\inst{3}
          }

   \offprints{a.chrysostomou@star.herts.ac.uk \\ Table 2 is only available in electronic form at the CDS via anonymous ftp to cdsarc.u-strasbg.fr (130.79.128.5) or via http://cdsweb.u-strasbg.fr/cgi-bin/qcat?J/A+A/}

   \institute{Dept of Physical Sciences, University of Hertfordshire, Hatfield, HERTS, AL10 9AB, UK\\
              \email{a.chrysostomou,d.aitken,j.hough@star.herts.ac.uk}
         \and
             Joint Astronomy Centre, 660 N. A'ohoku Place, Hilo, HI 96720, USA\\
             \email{t.jenness,c.davis@jach.hawaii.edu}
	\and
	     National Astronomical Observatory of Japan, Osawa, Mitaka, Tokyo 181-8588, Japan\\
	     \email{tamuramt@cc.nao.ac.jp}
             }

   \date{Submitted: September 11, 2001 ; Accepted : January 29, 2002 }

   \abstract{ We present 850~$\mu$m imaging polarimetry of the W51A
   massive star forming region performed with SCUBA on the JCMT. From
   the polarimetry we infer the column-averaged magnetic field
   direction, projected onto the plane of the sky. We find that the
   magnetic field geometry in the region is complicated. We compare
   the field geometry with 6~cm and CS~J=7-6 emission and determine
   that the magnetic field must be relatively weak and plays a passive
   role, allowing itself to be shaped by pressure forces and dynamics
   in the ionised and neutral gases. Comparisons are drawn between our
   data and 1.3~mm BIMA interferometric polarimetry data, from
   which we conclude that the magnetic field must increase in
   importance as we move to smaller scales and closer to sites of
   active star formation.  
   }

   \titlerunning{The magnetic field structure in W51A}
   \authorrunning{Chrysostomou et al.}

   \maketitle
%

\section{Introduction}

The W51A cloud consists of a complex of {\sc Hii} regions situated
approximately 7-8 kpc away. It is one of the most active and luminous
sites of massive star formation in the Galaxy with all the associated
signatures: masers (Lepp\"anen et al
\cite{Leppanen98}; Zhang \& Ho \cite{Zhang95}; Genzel \& Downes
\cite{Genzel77}), {\sc Hii} regions (Scott \cite{Scott78}), 
bright \& dusty cores (Jaffe et al. \cite{Jaffe84}; Genzel et
al. \cite{Genzel82}) and nearby supernova remnants (Koo \& Moon
\cite{Koo97a,Koo97b}). Radio maps show a number of 
{\sc Hii} regions, the most luminous being G49.5-0.4 (see
e.g. Mehringer \cite{Mehringer94}) which itself is dominated by the
sources W51d and W51e. The former is directly associated with its
infrared counterpart IRS2 (Goldader \& Wynn-Williams
\cite{Goldader94}) and while W51e and the infrared source 
IRS1 are also closely associated, this section of the cloud is more
complex.  A cluster of compact {\sc Hii} regions are found to the east
of W51e/IRS1, whose radio emission is dominated by the sources W51e2
and e1. The submillimetre emission is more closely associated with
this cluster rather than with IRS1 and the peak of emission correlates
with the position of W51e2 (Jaffe et al. \cite{Jaffe84}), suggesting
that W51e is a more evolved source while W51e1 and e2 are only now
undergoing star formation. This is corroborated by the detection of
infalling gas associated with W51e2 by Zhang \& Ho (\cite{Zhang97}).

Here, we present 850~$\mu$m polarimetry of the W51A region revealing
the magnetic field structure within it. There are a number of methods
which allow for mapping the magnetic field geometry, and each has its
own associated problems. By far the most promising method is through
polarimetry of emission from dust grains which have been aligned (it
is necessarily assumed) by interaction with the local magnetic
field. Previous polarimetric observations of W51A show very low
polarisation (Kane et al \cite{Kane93}) consistent with a non-detection. 
However, these measurements were made with large beams ($\sim$
30\arcsec) within which a significant degree of depolarisation can
occur. For instance, Dotson et al. (\cite{Dotson00}) present
far-infrared polarimetry for W51A at 100 $\mu$m taken with the
KAO. Their results show low polarisation at the cores and higher
degrees of polarisation in regions with more extended emission. Higher
resolution and more sensitive measurements are clearly necessary to
detect significant polarisation and to determine any structure in the
polarisation pattern (e.g. see Lai et al. \cite{Lai01} and Momose et
al. \cite{Momose01}). In their mid-infrared spectropolarimetric atlas
of young stellar objects, Smith et al. (\cite{Smith00}) measure 6\%
and 3\% polarisation towards W51d in absorptive and emissive
components, respectively, measurements made within 5\arcsec~beams.

The importance of magnetic fields in star formation has become
increasingly more apparent in recent years as computer models have
become more sophisticated (e.g. Ostriker et al. \cite{Ostriker01}). 
Magnetic fields are believed to play an important regulatory role
during cloud collapse and accretion onto the protostar, as well as
providing the driving and collimation mechanism for outflows.
However, in order to comprehend the magnetic field's role properly, an
empirical understanding of its behaviour is necessary.

In this paper, we present 850~$\mu$m polarimetry not only of the dense
cores in W51A but also of the surrounding more diffuse emission
between these cores. In this fashion we investigate the relation
between the magnetic field at the sites of star formation and the
cold, expansive envelopes that surround them. By comparing this
information with other studies we may also hope to understand what
shapes the magnetic field. For instance, are the kinematics of gas and
dust, as governed by gravitational collapse and the pressures within
{\sc Hii} regions, dominant, or is the energy density in the magnetic
field of sufficient magnitude to control and regulate the dynamics?


\section{Observations and Reduction}

The data for W51A were observed on the night 6 October 2000 on the
James Clerk Maxwell Telescope (JCMT) situated close to the summit of
Mauna Kea, Hawaii. The observations were made with the SCUBA camera
(Holland et al. \cite{scuba}) together with a polarimetry module in
the beam mounted on the entrance window of SCUBA. The polarimetry
optics consist of a rotating quartz half-wave retarder ahead of a
fixed wire-grid analyser (see Greaves et al. \cite{scupol} for a full
description of the polarimeter). To produce a fully sampled map at
850~$\mu$m, 16 offset positions of the array are needed on the sky
(Holland et al.). The resultant image is referred to as a ``jiggle
map''. For these observations of W51A an east-west chop throw of
140\arcsec~at 7.8~Hz was used. Complete jiggle maps were taken at
specific positions of the half-wave retarder, each separated by
22.5\degr. This therefore requires 16 jiggle maps to complete one
cycle of the retarder. Six such cycles were observed towards W51A.

The SURF (SCUBA User Reduction Facility; Jenness \& Lightfoot
\cite{surf}) data reduction package was used to flatfield, rebin and
sky noise correct the SCUBA data. Excessively noisy bolometers were
identified from noise measurements conducted on the observing night
and ``switched-off'' so that they take no further part in the data
reduction. Sky subtraction with SURF is used for eliminating sky noise
in SCUBA data (Jenness et al. \cite{jenness98}). Sky
noise can be introduced, for example, by variations in the sky
transparancy between the separate positions of the secondary, which
make up the jiggle pattern, in a single exposure. To correct for this
noise bolometers devoid of any significant flux which can be
attributed to point sources, or to extended emission, are chosen. The
signal from each jiggle position in these bolometers is then used to
calculate and correct for any temporal variation in the sky signal. To
preserve the total flux in the map, the mean background signal which
is removed in this correction is then added back. Inspection of the
final image showed that the weakest signal is an order of magnitude
fainter than that of the diffuse emission. This weak signal being
$\sim 2\sigma$ above the sky noise, suggests that conditions were
stable enough over the timescale of the chop cycle that sky background
was efficiently removed by the chop. The very small, residual sky
emission would have to be quite highly polarised to have any effect on
our results and conclusions.  With the sky noise removed, the data are
then rebinned onto a 3.09\arcsec ~pixel grid, representing half the
spacing of the jiggle pattern, to form the final images.

Before rebinning the images and calculating the polarisation, the main
lobe instrumental polarisation (IP) is removed from each bolometer
using a look-up table of elevation dependent IP values. This software
also corrects for the parallactic angle at the time of observations by
altering the value of the effective retarder position on the
sky. Following Greaves et al. (\cite{scupolcomm}) and Matthews et
al. (\cite{Matthews01}), contamination due to sideband lobe
instrumental polarisation is estimated to be negligible (i.e. within
errors) for our data.

The $I$, $Q$ and $U$ Stokes parameters were computed for each cycle of
the retarder using the POLPACK polarimetry data reduction software
(Berry \& Gledhill \cite{polpack}). For each cycle, images taken at
equivalent positions of the retarder are averaged together
(e.g. positions at 0\degr, 90\degr, 180\degr ~and 270\degr
~equivalently measure the same angle on the sky). The result is a
stack of four images where each pixel in the stack samples the
modulated signal at angles close to 0\degr, 22.5\degr, 45\degr ~and
67.5\degr ~to which the following curve is fitted to extract the
Stokes parameters (Sparks \& Axon \cite{sparks99}):

\begin{equation}
I' = \frac{t}{2} (I + \epsilon [Q \cos 2\phi + U \sin 2\phi])
\end{equation}

\noindent
where $I'$ is the expected intensity, $t$ is the throughput of the
wire-grid analyser, $\epsilon$ is the polarising efficiency of the
analyser, and $\phi$ is the effective retarder position angle after
correction for the parallactic angle.

After combining the data from the six cycles, the Stokes parameters
are binned into $3\times3$ pixel bins (equivalent to
9.3\arcsec). Finally, the polarisation is debiassed and thresholded so
that only polarisation vectors associated with positive flux ($I > 0$)
and small polarisation error ($\delta P < 0.5\%$) are included in the
final data set. The bulk of our polarisation vectors lie in the 1--3\%
range which, when the maximum polarisation error of $\delta P = 0.5\%$
is taken into account, translates to an uncertainty for
the position angles of $\Delta \theta \sim 28.6\degr/\sigma_P \sim
15-5\degr$.

Finally, flux calibration was performed by multiplying the data
by a ``flux calibration factor'' (FCF) of 414 Jy/beam/Volt. This FCF
has been shown to be repeatably accurate to $\sim$~10\% for all SCUBA
calibrations (Jenness et al. 2002). The value used here has been
modified to account for the loss of throughput due to the polarimeter
(Greaves et al. 2002).

\begin{table}
  \caption[]{Brightest {\sc Hii} regions and their positions in the W51A cloud. Positions obtained from the SIMBAD database.}
  \label{sources}
	\begin{tabular}[t]{lll}
		Object & RA (J2000) & Dec(J2000) \\
		\hline 
		W51d (IRS2) & 19 23 39.9 & +14 31 08 \\
		W51e (IRS1) & 19 23 41.9 & +14 30 36 \\
		W51e1       & 19 23 43.8 & +14 30 25.9 \\
		W51e2       & 19 23 43.9 & +14 30 34.8 \\

	  \end{tabular}
\end{table}

\section{Results and Discussion}

\subsection{Submillimetre polarimetry}

\begin{figure*}
	\includegraphics[width=\textwidth]{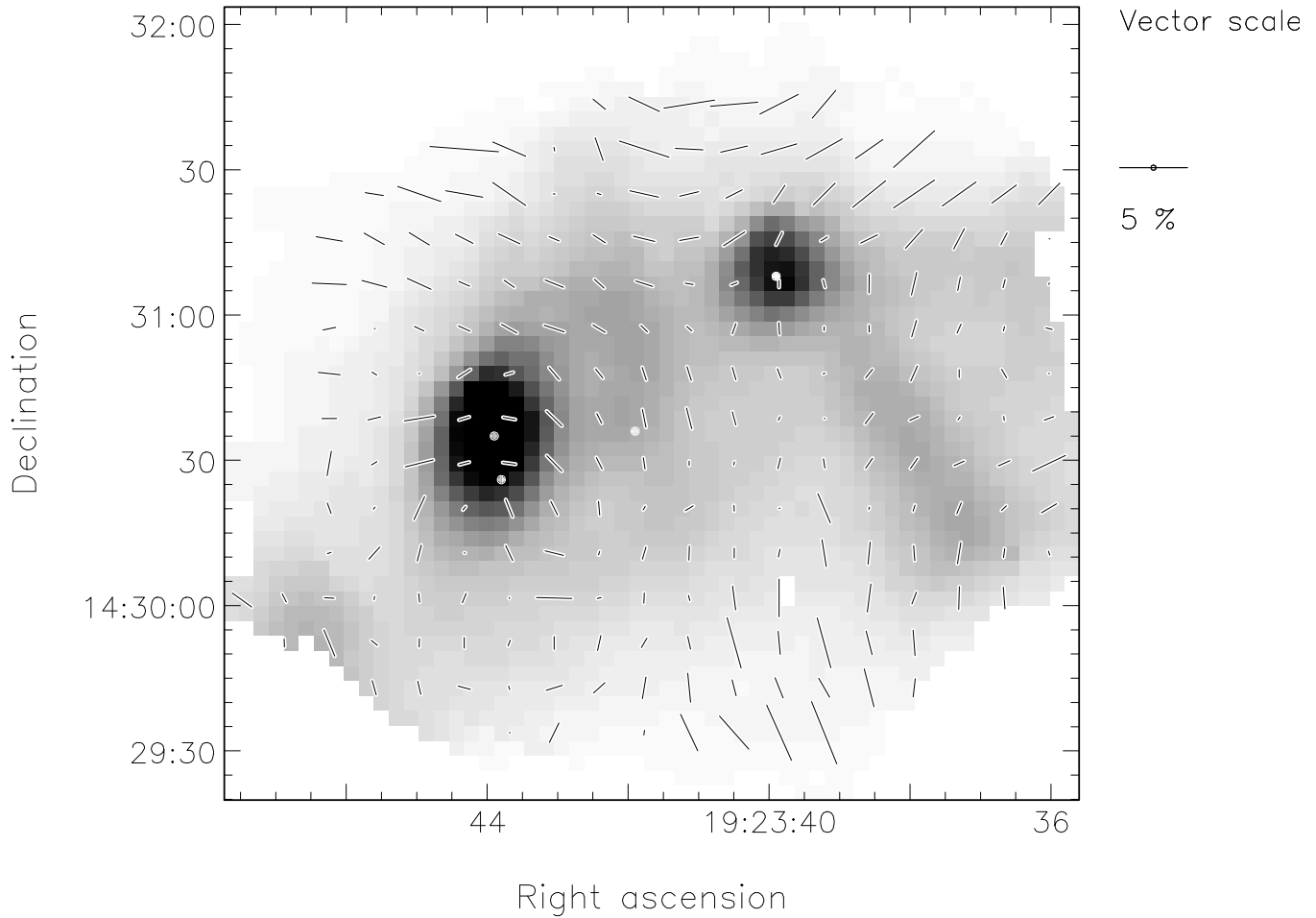}

	\caption{SCUBA polarimetry of the W51A region. The intensity
	image is taken from our data and is plotted on a linear
	scale. The source to the east of the field is the
	submillimetre core associated with W51e1 and e2, and the
	source to the west is W51d. The polarisation vectors have
	been rotated by 90\degr~and plotted such that their position
	angles delineate the direction of the magnetic field averaged
	along the column and projected onto the plane of the sky. The
	white dots represent the positions of the brightest {\sc Hii}
	regions in the field (see Table \ref{sources}). They are, from
	right-to-left: W51d (IRS2), W51e (IRS1), W51e1 and W51e2 (e1
	is south of e2). Right ascension and Declination are at epoch
	2000.}

	\label{w51pol}
\end{figure*}

The final results are tabulated in Table 2 and shown in
Fig. \ref{w51pol}. The greyscale shows the submillimetre emission
which is dominated by two bright sources. To the east is the core
which was mapped by Jaffe et al. (\cite{Jaffe84}) at 400~$\mu$m and is
associated with the W51e1 and e2 compact {\sc Hii} regions (Mehringer
\cite{Mehringer94}). The submillimetre core to the north-west is
associated with the W51d {\sc Hii} region and with the IR source IRS2
(Goldader \& Wynn-Williams \cite{Goldader94}). Both cores are known
sites of maser activity (Genzel \& Downes \cite{Genzel77}). Fainter
and more extended emission is also seen, occupying the region between
the two cores, and an extended arm of emission is seen to reach out to
the south-west from IRS2.

Fig. \ref{w51pol} shows polarisation vectors overlaid onto the
submillimetre emission. These vectors have been rotated by 90\degr ~to
depict the column-averaged direction of the magnetic field projected
onto the plane of the sky. This field direction towards the W51e1/e2
core is at a position angle of $\sim$~100--110\degr, consistent with
recent interferometer polarimetry at 1.3~mm with BIMA (Lai et
al. \cite{Lai01}). In an atlas of mid-infrared spectropolarimetry of
young stellar objects, Smith et al. (\cite{Smith00}) observed
W51d(IRS2) and deconvolved emissive and absorptive components to the
spectropolarimetry. The emissive ({\bf E}-vector) component was at
$36\degr \pm 3\degr$ implying a field direction in the hotter regions
at or near the {\sc Hii} region (there is a clear [Ne{\sc II}] line in
the spectrum) where the 10~$\mu$m emission peaks of 126\degr. The
absorptive component of the {\bf E}-vector, and therefore equivalent
to the {\bf B}-vector, is at a position angle of 136\degr, a
difference of $\sim$~40\degr when compared to our measurement of
$\sim$~0\degr. Being absorptive, the dust responsible for the
mid-infrared polarisation is also probably largely responsible for the
submillimetre emission. This difference in the inferred magnetic field
directions is most likely due to structure in the magnetic field being
sampled to different optical depths at the two wavelengths (and the
different beam sizes to some extent).

\begin{figure}
	\includegraphics[width=8.5cm]{H3147F2A.eps}
	\includegraphics[width=8.5cm]{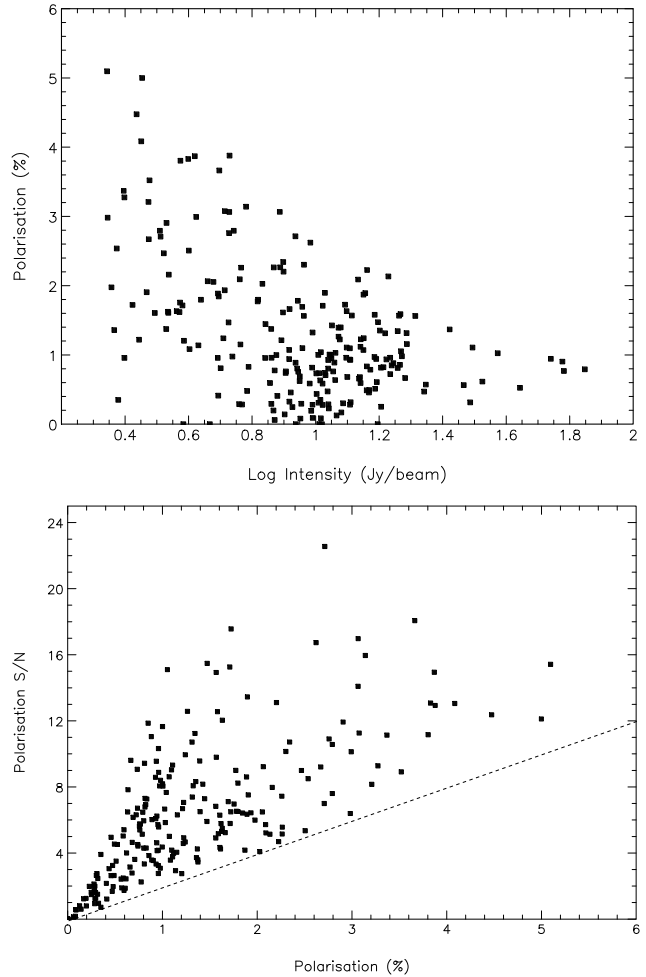}

	\caption{Scatter and histogram plots of the polarimetry
	presented in Fig. \ref{w51pol}. The top panel shows the
	degree of polarisation variation with the logarithm of
	intensity. The bottom panel shows the variation of
	$\sigma_P$, with polarisation. No points exist below the
	dotted line due to the data-clipping of points with
	polarisation errors $\delta P > 0.5\%$ (see Sect. 2). }

	\label{scatter}

\end{figure}

Scatter plots of the polarisation data (Fig. \ref{scatter}) show
that the degree of polarisation has a maximum of $\sim$ 5\% and also
that the polarisation signal-to-noise, $\sigma_P$, ranges up to a
maximum value $\sim$~20. The top panel of Fig. \ref{scatter} shows
that across the region the polarisation and intensity are generally
anti-correlated, a trend seen by others (e.g. Matthews et
al. \cite{Matthews01}; Davis et al. \cite{Davis00}). A number of
physical mechanisms have been attributed to explain this effect. A
decrease in grain alignment efficiency as a function of increasing
depth into a cloud is often cited as the cause (e.g. Lazarian et
al. \cite{Lazarian97}). This is because most known mechanisms of grain
alignment are believed to decrease in efficiency at the physical
conditions common to cold, dark clouds. On the other hand, small scale
field structure induced by field lines being dragged and twisted under
the influence of gravitational collapse, for instance, are capable of
producing the {\em depolarising} effect if the telescope beam is
larger than the scale lengths over which these effects occur
(e.g. Aitken et al. \cite{Aitken97}). Zhang et al. (\cite{Zhang98})
present evidence which suggest that cores within IRS1 are indeed
collapsing and rotating. However, as a caveat to this it should
be noted that this trend of low degrees of polarisation for high
submillimetre intensities is not universal as can be seen in recent
submillimetre polarimetric imaging of the NGC7538 region (Momose et
al. \cite{Momose01}) where the degree of polarisation remains
relatively high towards the cores.

It is also worth noting that polarisation nulls can be seen to
the north-east and south-west of W51e1/e2, at positions lying almost
perpendicular to the field direction defined at the core. These null
vectors lie at symmetric positions, that is, they are both equispaced
and on the same line through the source centre. The data have been
checked and there is no apparant systematic reason for these vectors
to be lowly polarised apart from the fact that the polarimetry at
these positions has low signal-to-noise. A physical reason may be that
at these positions the field becomes wound up such that the
polarisation drops to very low values ($P <$ 0.5\%) and the polarised
intensity is averaged out within the relatively large JCMT beam. We
choose to briefly discuss these features here because simulations do
show that such polarisation patterns are possible for pinched and
twisted field morphologies (Aitken et al. \cite{Aitken01}) as well as
for helical fields (Fiege \& Pudritz \cite{Fiege00}), although the
latter deals with linear structures and is not really applicable to
discrete sources. Furthermore, such `submillimetre polarisation nulls'
either side of a bright core have been previously observed towards
NGC7538 IRS1 (Momose et al. \cite{Momose01}). In both of these cases,
the field pattern appears very much like the familiar polarisation
disks seen in scattered light, at near-infrared and optical
wavelengths, around sources with optically thick disks (e.g. Lucas \&
Roche \cite{Lucas98}). Higher spatial resolution observations would
clearly help us to understand the physical basis behind these
polarisation nulls, as we note that there are nulls present in
our data at other positions in the region.

An immediately striking feature of the polarimetry as a whole is that
the magnetic field threading the region does not appear uniform in
direction or degree of polarisation. It is interesting to note that
the polarisation vectors which are positioned central and to the south
in Fig. \ref{w51pol} lie closely aligned to the projections of the
Galactic plane and the large scale Galactic magnetic field onto the
plane of the sky (PA $\sim 20$\degr; Matthewson \& Ford
\cite{Mathewson71}). In their study of water masers in the core of
W51e2, Lepp{\"a}nen et al. (\cite{Leppanen98}) found the masers' proper
motions also aligned in this direction. Muson \& Liszt
(\cite{Mufson79}) find a large scale stream of molecular gas at this
same postion angle passing through the centre of the W51A
region. Their CO data shows a complex of velocities with the bulk
motion at 58~km~s$^{-1}$, the streamer velocity at 70~km~s$^{-1}$ and
cold absorbing gas at $\sim$ 66~km~s$^{-1}$. Koo \& Moon
(\cite{Koo97b}) determined that $^{13}$CO emission is associated with
the streamer.

Clearly, this position angle of $\sim$ 20\degr~ bears some
significance for this region. Those vectors which closely follow this
position angle, may be tracing the large-scale Galactic magnetic
field which follows the Galactic plane and may also be dragged and
stretched by material in the streamer. The field then becomes
disrupted when it encounters the dense, star-forming cores. Towards
W51e1/e2 and W51d the field appears perpendicular to this global
direction, suggesting that some degree of field dragging and
winding-up must have occured during the initial collapse of these
cores.

The arm of submillimetre emission which extends to the south-west of
W51d also has the field associated with it perpendicular to the
Galactic plane and the arm itself. It appears that the collapse to
form this elongated structure has occured along the magnetic field
lines, but this does not seem probable if those field lines are
connected to field lines to the north and to the south of the
structure, where the field apparantly joins the Galactic Plane. It 
could be that the gravitational collapse has dragged the field
lines into this pattern.

\begin{figure*}
	\includegraphics[width=\textwidth]{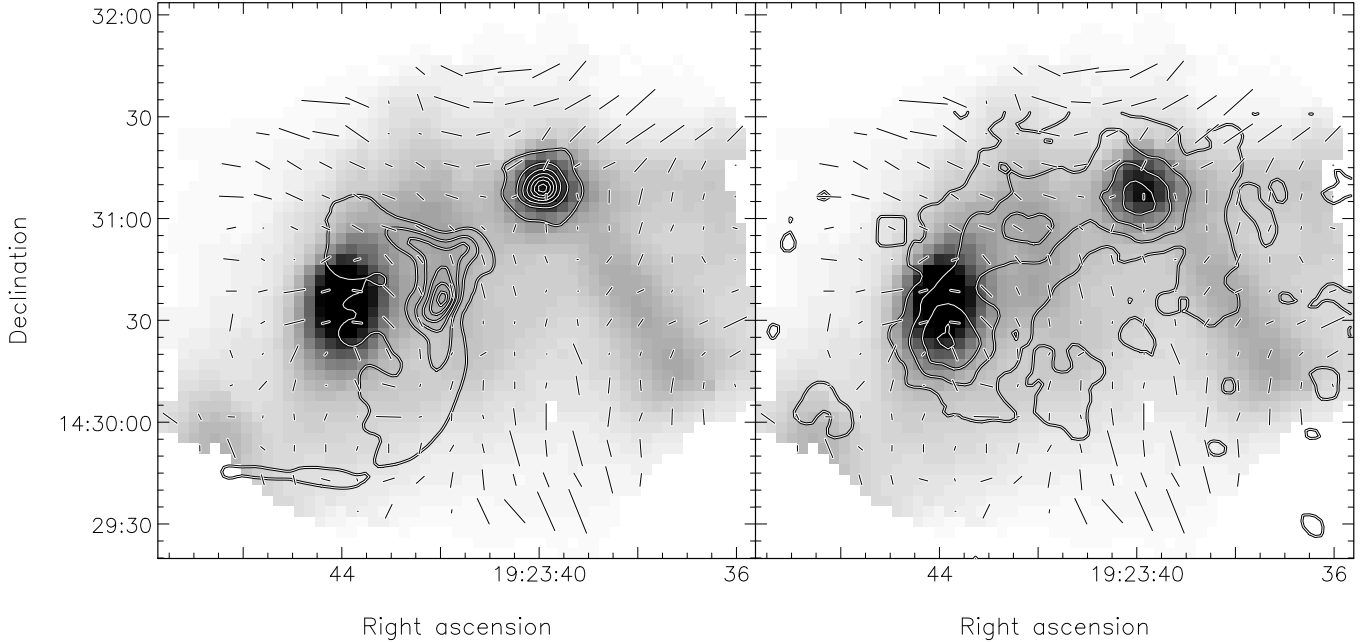}

	\caption{Submillimetre polarimetry image of Fig.
	\ref{w51pol} overlaid with contours of 6cm emission (left
	panel; Mehringer \cite{Mehringer94}, courtesy of ADIL) and CS
	J=7-6 emission (right panel). The presence of the W51e {\sc
	Hii} region associated with IRS1 seems to be influencing the
	magnetic field. The CS emission (integrated between 59 and
	63~km~s$^{-1}$) reveals a bridge of dense gas between W51e1/e2
	and W51d (IRS2). Once again, the magnetic field appears to be
	influenced by this structure, especially the northern portion
	of the bridge. Right ascension and Declination are at epoch
	2000. }

	\label{twoplot}
\end{figure*}

\subsection{Comparison with 6~cm and CS~J=7-6 emission}

We have obtained a 6~cm emission map of the W51A region, downloaded
from the {\em NCSA Astronomical Data Image Library}. The data were
originally acquired at the VLA and published by Mehringer
(\cite{Mehringer94}). Also, raw data from a CS~J=7-6 raster map of the
W51A region were retrieved from the JCMT archive (operated by the
CADC) and reduced. Fig. \ref{twoplot} shows contour plots of the
6~cm and CS emissions overlaid onto our submillimetre measurements.

The 6~cm map traces free-free emission from ionised gas and the
eastern peak, W51e, coincides with the IRS1 source seen in the
infrared. The extended structure of the 6~cm emission seems to
correlate with the magnetic field direction quite well. The extended
emission as a whole gives the impression of an envelope around the
western edge of the submillimetre core and, from the polarisation
pattern, the magnetic field appears to be influenced by the presence
of this ionised gas. There is an extension to the north-east of W51e
which follows the magnetic field lines and may form part of the
molecular streamer seen by Mufson \& Liszt (\cite{Mufson79}).  In all,
the 6~cm emission possesses a degree of curvature which may be a
consequence of the expansion of the {\sc Hii} region being impeded by
the dense submillimetre core. If one follows the inferred magnetic
field direction from the north-east through to the south-west, then
one finds that around the location of W51e the field is indeed curved.

Being a density tracer, the integrated intensity map of the CS~J=7-6
emission looks very similar to the submillimetre emission. However,
the contour map presented in Fig. \ref{twoplot} shows the integrated
flux between 59 and 63~km~s$^{-1}$, and at these velocities, a bridge
of material between the two cores becomes apparant. Channel maps of
the CS data show that there exists a small velocity gradient of $\sim$
3~km~s$^{-1}$ across the bridge.  This suggests a flow along the
bridge and between the two cores, although this evidence may be better
interpreted in terms of a rotating structure. Proper dynamical
modelling, beyond the aims and scope of this paper, is necessary for a
better understanding of this structure.

The relation of the magnetic field with the material responsible for
the CS J=7-6 emission is not as clear as for the 6~cm emisssion. The
most intriguing feature in the CS emission is clearly the bridge
structure between the two cores, a feature not pronounced in the
850~$\mu$m emission. From the W51e1/e2 core, the bridge extends to the
north-west. Here the magnetic field seems to lie perpendicular to the
bridge, although it could easily be more influenced by the molecular
streamer in the Galactic Plane (see Sect. 3.1) or the nearby W51e
{\sc Hii} region. Further along, the bridge then curves around to be
more latitudinal; the magnetic field fans out and is now parallel with
the bridge. If there really is material flow along this structure,
then it may be that the magnetic field is being pulled and stretched
along the bridge at this position.

\subsection{An upper limit to the magnetic field strength}

As well as the morphology of the magnetic field, it is also important
to try and obtain estimates of the magnetic field strength. Methods
that have been used in the literature have included the Zeeman effect
(e.g. Crutcher \cite{Crutcher99}), which gives an indication of the
strength of the field component along the line of sight, and also the
method of Chandrasekhar \& Fermi (\cite{CF}) which gives the strength
of the field component along the plane of the sky through measurement
of the dispersion of polarisation position angles. The latter method
is applicable to our data set but would not return sensible results
due to our large beam and also because the large dispersion of
position angles we find across our field renders this method futile
(see Sect. 6 of Ostriker et al. \cite{Ostriker01}). However, we are
able to place quite reliable upper limits to the magnetic field
strength through consideration of the Virial Theorem. For a
self-gravitating molecular cloud with negligible external pressure the
virial equilibrium of that cloud can be written as (McKee et
al. \cite{McKee93}; Glenn et al. \cite{Glenn99}):

\begin{equation}
\left| \frac{-3GM^2}{5R} \right| = 2\times0.27M\Delta v^2 + 0.1B^2R^3
\end{equation}

For W51e2 we use a mass of 800~M$_{\sun}$ (Zhang et
al. \cite{Zhang98}), $\Delta v$~$\sim$~1~km~s$^{-1}$ (Young et al.
\cite{Young98}), and a radius for the cloud of
$R~\sim~3\times10^{15}$~m, measured from the 1.3~mm map of Lai et
al. (\cite{Lai01}), to obtain a value of $B_{\rm{up}}~\sim~3.5$~mG. For W51d we
use a mass of 400~M$_{\sun}$ (Zhang et al. \cite{Zhang98}), $\Delta
v$~$\sim$~1~km~s$^{-1}$ again, and a radius for the cloud of
$R~\sim~5\times10^{15}$~m, measured from the 2~mm map of Zhang et
al. (\cite{Zhang98}), to obtain a value of $B_{\rm{up}}~\sim~0.6$~mG.

Since we know that these clouds are in the process of collapse and
forming stars, these values form useful upper limits to the value of
the magnetic field strength. The value for W51e2 is consistent with
values obtained using the method of Chandrasekhar \& Fermi
($\sim$~1~mG; Lai et al.). Crutcher (\cite{Crutcher99}) presented a
statistical analysis of Zeeman measurements from a number of molecular
clouds (but not W51) and found line-of-sight field strengths of the
order of 0.5~mG for clouds with densities of 10$^{5-6}$~cm$^{-3}$.

\section{The Magnetic Field in W51A: Just How Important Is It?}

In Sect. 3 above, we describe how the magnetic field follows the
curvature of the extended 6~cm emission and is also influenced by the
bridge of CS~J=7-6 emission between the two submillimetre cores. The
apparant curvature of the field and the manner in which the field
aligns with the velocity gradient in the CS bridge can be interpreted
as evidence that the magnetic field is being shaped by the pressure
and dynamics of the gas. Similar magnetic field behaviour has been
seen in recent SCUBA polarimetry of NGC 7538 (Momose et
al. \cite{Momose01}) where the blue- and red-shifted outflow lobes
from IRS11 are observed to have quite a high degree of curvature. The
magnetic field inferred from the polarimetry follows the same
curvature. However, one would expect that if the field were of
sufficient strength, magnetic tension in the field lines would resist
such action and return them back into an ordered and linear
configuration.

Towards the core of the W51e1/e2 region we find that the magnetic
field direction is oriented roughly perpendicular to the long axis of
the extended envelope. This is clearly suggestive that the material
collapsed along the field lines and then fragmented to form the cores
now present. This interpretation is advocated by Lai et
al. (\cite{Lai01}) who found a very uniform field direction across the
two cores (with the benefit of 3\arcsec~resolution from their
interferometry data) and interpreted this to mean that the magnetic
field plays a dominant role. However, rotational motions have been
detected towards the collapsing cores of W51e1 and e2 (Zhang et
al. \cite{Zhang98}), and their rotational axes are between
40--90\degr~ from that of the magnetic field. This misalignment is
evidence that the magnetic field has had no influence in determining
the rotational axes for these clumps, evidence that their
fragmentation and collapse from the common envelope has been guided
according to physical parameters other than the magnetic field.

At the small scales studied by Lai et al., their reasoning may well
hold true but the evidence presented here suggests that the magnetic
field in W51A does not play a dominant role in determining where gas
flows. Firstly, across the 2\arcmin~field-of-view we find a large
dispersion of polarisation position angles (see Fig.
\ref{w51pol}). Recent MHD modelling of a turbulent molecular cloud
(Ostriker et al. \cite{Ostriker01}) has shown that, as
expected, clouds with weak fields ($\beta~\sim$~1, where $\beta$ is
the ratio of the squares of the sound and Alfv\'en speeds and
parameterises the relative strength of the magnetic field; see
Eq. \ref{beta}) produce polarisation maps with vectors which are
less organised and ordered than those with stronger fields
($\beta~\sim$~0.01). This is consistent with what we are observing in
W51A (see Fig. 23 in Ostriker et al.).

We may analyse this further by investigating the dimensionless
parameter, $\beta$. This parameter is defined as (Ostriker
et al. \cite{Ostriker01}):

\begin{equation}
\beta = \frac{c_{\rm{s}}^2}{v_{\rm{A}}^2} = \frac{c_{\rm{s}}^2}{B_0^2 / 4\pi\rho}
\label{beta}
\end{equation}

\noindent
where $c_{\rm{s}}$ is the sound speed, $v_{\rm{A}}$ is the Alfv\'{e}n
velocity, $B_0$ is the magnetic field strength and $\rho$ is the mass
density.  Using a sound speed of $c_{\rm{s}}^2 = 3kT/m$, we can
rewrite the $\beta$ parameter as the ratio of the gas to magnetic
pressures:

\begin{equation}
\beta = 12\pi \frac{nkT}{B_0^2}
\end{equation}

\noindent
where $n$ is the particle number density, $T$ is the temperature and
$k$ is Boltzmann's constant.  Taking estimates of density and
temperature from the literature we find that
$n~\sim~4\times10^5~-~10^6$~cm$^{-3}$ and $T~\sim~20-35$~K (see
e.g. Lai et al. \cite{Lai01} and Jaffe et al. \cite{Jaffe84}). We can
now write the $\beta$ parameter specifically for the W51A region as:

\begin{equation}
\beta \sim 10^{-8}/{B_0^2}
\end{equation}

In their simulations, Ostriker et al. defined a `strong field'
condition when $\beta~=~0.01$ which produced well correlated and
ordered polarisation maps. This translates to a field strength of
$B_0~\sim$~1~mG in W51A. Conversely, their weak-field case has
$\beta~=~1$ which translates to a field strength of $B_0~\sim$~0.1~mG.

We therefore find a consistency between our data, which infers a weak
field on large scales, and that of Lai et al., who infer a strong
field on smaller scales. The magnetic field has an important role to
play in the W51A region as one approaches the dense cores where star
formation is active. However, as one moves to the more expansive
envelope and to regions between star forming cores, the magnetic field
appears weaker and plays a more passive role, allowing itself to be
shaped by the pressure forces and dynamics of the gas.


\section{Conclusions}
	
We present 850~$\mu$m polarimetry of the massive star forming region
in W51A. The presented data has relatively high polarimetric
signal-to-noise, which allows us to infer a magnetic field structure
which is not uniform across our 2\arcmin~field-of-view.

We have compared our data with a VLA 6~cm emission map as well as with
a CS~J=7-6 map obtained from the JCMT archive. The comparison shows a
correlation between the magnetic field geometry and the ionised and
molecular gas. This indicates that the magnetic field is shaped by the
gas dynamics and not the reverse, and therefore has a relatively weak
magnetic field strength (parameterised by $\beta~\sim~1$).

However, Lai et al. (\cite{Lai01}) have found from their 1.3~mm
interferometer polarimetry that at smaller scales and towards the star
forming cores the field lines are regular and ordered, implying that
the magnetic field is comparatively strong ($\beta~\sim~0.01$) and
plays a significant role in determining the gas dynamics -- although
evidence suggests that further fragmentation and collapse probably
occurred independent of the magnetic field. Nevertheless, when one
looks at larger scales and further comparisons are made between the
global field morphology and the gas dynamics, this picture breaks
down.

Our data has shown how important it is to attempt to correlate
magnetic field structures with the physical conditions of the (ionised
and neutral) gas. In this way, one can develop a better understanding
of the role magnetic fields play in star formation.

\begin{acknowledgements}
The JCMT is operated by the Joint Astronomy Centre in Hilo, Hawaii on
behalf of the parent organizations Particle Physics and Astronomy
Research Council in the United Kingdom, the National Research Council
of Canada and The Netherlands Organization for Scientific
Research. The Canadian Astronomy Data Centre (CADC), is operated by
the Herzberg Institute of Astrophysics, National Research Council of
Canada. This research has made use of the SIMBAD database, operated at
CDS, Strasbourg, France. The authors would like to thank the NCSA
Astronomy Digital Image Library (ADIL) for providing images for this
article. We acknowledge the support through software and data analysis
facilities provided by the Starlink Project which is run by CCLRC on
behalf of PPARC.

The authors would like to thank the referee, Dan Clemens, for
carefully reading our original manuscript and providing us with useful
comments which improved on it.

\end{acknowledgements}


\end{document}